\documentclass[prd,aps,twocolumn,a4paper,nofootinbib]{revtex4-1}

\newif\ifusesec
\usesectrue % or
\usepackage{graphicx,psfrag}
\usepackage{mathrsfs}
\usepackage{amsmath,amsfonts,amssymb}
\usepackage{multirow}
\usepackage{comment}

\DeclareSymbolFontAlphabet{\mathrsfs}{rsfs}
\DeclareMathAlphabet\mathbfcal{OMS}{cmsy}{b}{n}

\newcommand{\be}{\begin{equation}}
\newcommand{\ee}{\end{equation}}
\newcommand{\bea}{\begin{eqnarray}}
\newcommand{\eea}{\end{eqnarray}}
\newcommand{\bel}{\begin{align}}
\newcommand{\eel}{\end{align}}

\def\i{{\rm i}}
\def\h{{\bar{h}}}

\def\GMc2{G M_{\odot} c^{-2}}

\def\ii{{\rm i}}

\DeclareSymbolFontAlphabet{\mathrsfs}{rsfs}
\DeclareMathAlphabet{\mathcal}{OMS}{cmsy}{m}{n}

\usepackage{color}

\definecolor{cyan}{rgb}{0,0.9,0.9}
\definecolor{orange}{rgb}{0.9,0.5,0}
\definecolor{magenta}{rgb}{1,0,1}
\definecolor{purple}{rgb}{0.8,0.4,0.8}
\definecolor{gray}{rgb}{0.8242,0.8242,0.8242}

\begin{document}

\title{A new analytic representation of the ringdown waveform\\
        of coalescing spinning black hole binaries}

\author{Thibault \surname{Damour}}
\affiliation{Institut des Hautes Etudes Scientifiques, 91440 Bures-sur-Yvette, France}
\author{Alessandro \surname{Nagar}}
\affiliation{Institut des Hautes Etudes Scientifiques, 91440 Bures-sur-Yvette, France}

\date{\today}

\begin{abstract}
We propose a new way of analyzing, and analytically representing, the ringdown part of the gravitational 
wave signal emitted by coalescing black hole binaries.
By contrast with the usual {\it linear} decomposition of the multipolar complex waveform $h(t)$ in a 
sum of quasi-normal modes, our procedure relies on a {\it multiplicative} decomposition of $h(t)$
as the product of the fundamental quasi-normal mode with a remaining time-dependent complex factor
whose amplitude and phase are separately fitted.
As an illustrative example, we apply our analysis and fitting procedure to the ringdown part of a sample 
of sixteen $\ell=m=2$ equal-mass, spinning, nonprecessing, numerical waveforms computed with the SP$_{\rm E}$C code,
now publicly available in the SXS catalogue. Our approach yields an efficient and accurate way to represent 
the ringdown waveform, thereby offering a new way to complete the analytical effective-one-body inspiral-plus-plunge 
waveform.

\end{abstract}


\maketitle

\section{Introduction}

The numerical-relativity (NR) completion of the effective-one-body (EOB) 
approach~\cite{Buonanno:1998gg,Buonanno:2000ef,Damour:2000we,Damour:2001tu,Buonanno:2005xu,Damour:2008gu,Damour:2009wj} 
(usually called EOBNR) is a NR-informed analytical method that aims at giving an accurate modelization of 
the gravitational dynamics and waveforms of coalescing relativistic binaries 
(i.e., black holes and neutron stars)~\cite{Buonanno:2007pf,Damour:2007yf,Damour:2007vq,Damour:2008te,Boyle:2008ge,Buonanno:2009qa,
Damour:2011fu,Pan:2011gk,Taracchini:2012ig,Damour:2012ky,Taracchini:2013rva,Pan:2013tva,Pan:2013rra,Damour:2013tla,Damour:2014afa}.
% to be used, in a computationally rather inexpensive way,
% to compute template waveforms for gravitational wave (GW) astronomy.
The EOB waveform for coalescing black-hole binaries is essentially made of the juxtaposition of 
two distinct parts: the inspiral-plus-plunge (or ``insplunge'') part (up to merger), and the subsequent
ringdown part (after merger). The insplunge waveform is analytically defined
by applying a sophisticated resummation procedure~\cite{Damour:2007xr,Damour:2008gu,Pan:2010hz} 
to the post-Newtonian-expanded waveform and dynamics~\cite{Blanchet:2013haa}. At merger, the insplunge waveform 
is {\it matched} to the ringdown part, defined, up to now, as a linear superposition of quasi-normal modes (QNMs) of the final black hole. 
The standard approach (initiated in Ref.~\cite{Buonanno:2000ef}) to compute this ringdown part is: 
(i) to identify the mass $M_{\rm BH}$ and angular momentum $J_{\rm BH}$ of the final black hole (either using the prediction 
of the EOB dynamics or using NR fitting formulas for these quantities~\cite{Hemberger:2013hsa,London:2014cma}); 
(ii) to use  $(M_{\rm BH}, J_{\rm BH})$ to compute a set of QNM frequencies~\cite{Berti:2005ys,berti:web}; and (iii) to build 
a linear superposition of QNMs with coefficients determined by imposing some matching (i.e., continuity) conditions 
between such a ringdown waveform and the EOB insplunge waveform at (iv) a certain {\it merger moment} $t=t_0$ determined 
from the EOB dynamics (e.g., as the peak of the EOB orbital frequency).
Such a procedure was found to work at a satisfactory level of robustness and accuracy in the nonspinning case 
(when complemented with some additional procedures, such as a matching over an interval 
around the time $t_0$)~\cite{Damour:2007xr,Damour:2007vq,Taracchini:2012ig,Damour:2013tla}.
However, the spinning case (as well as the case of higher modes in the nonspinning case~\cite{Pan:2011gk}) 
proved to be more challenging because the gravitational wave frequency of a matched QNM-ringdown signal was found 
to rise too quickly after merger~\cite{Taracchini:2012ig,Taracchini:2013rva}.
To overcome this difficulty, Refs.~\cite{Pan:2011gk,Taracchini:2012ig,Taracchini:2013rva} 
proposed to augment the analytical  ringdown signal by including, besides the real QNMs modes, 
some {\it pseudo-quasi-normal modes}, i.e., fictitious modes with frequencies phenomenologically chosen so as 
to bridge the gap between the final gravitational wave frequency of the insplunge EOB waveform measured 
at merger and the frequency of the fundamental QNM.

%The fact that the higher and positive the spin is, the closer the frequencies 
%and damping times of the first few mode become, made the straightforward matching procedure outlined above 
%completely unreliable~\cite{Pan:2009wj}. 

%=========================
% FIG. 1 Complex waveform vs time
%=========================
\begin{figure}[t]
\begin{center}
 \includegraphics[width=0.5\textwidth]{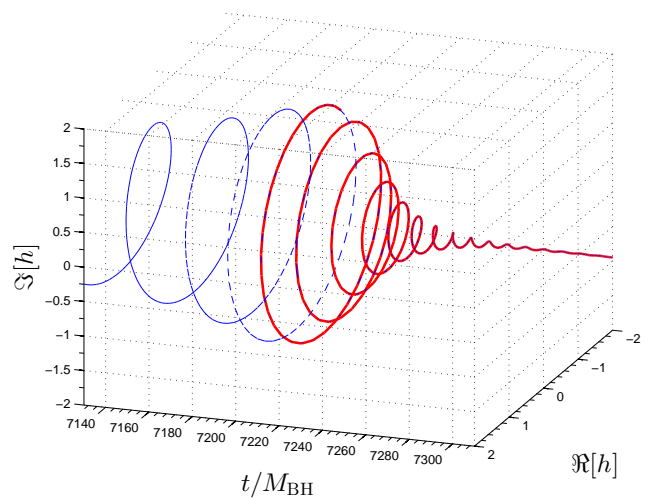}
    \caption{ \label{fig:wave_cmplx}Three-dimensional representation of the time evolution of the 
               $\nu$-rescaled {\it complex} strain (metric) waveform $h\equiv h_{22}/\nu$, 
    	       around merger, for dimensionless spin $\chi=0.97$. The thicker portion of the curve 
               (red online) highlights the ringdown part.} 
\end{center}
\end{figure}
%----------------------------------------

%=====================
% Table: QNMs values
%=====================
\begin{table*}[t]  
\caption{\label{tab:QNMs} Dimensionless ($M_{\rm BH}$-rescaled) complex frequencies of the first 
 three QNMs for three representative values of $\chi$.}
  \begin{center}
    \begin{ruledtabular}
\begin{tabular}[c]{lllll}
    $\chi$   & $\hat{a}_{\rm BH}$ & $\sigma_1=\alpha_1 + \ii\,\omega_1$ & $\sigma_2=\alpha_2 + \ii\,\omega_2$ & $\sigma_3=\alpha_3 + \ii\,\omega_3$ \\
\hline
    $-0.94905$  & 0.37567 &$0.0871184 +\ii\, 0.434580$   & $0.265632 + \ii\, 0.415053$  & $0.455361 + \ii\, 0.381143$\\
    $0    $  & 0.68703 &$0.0812684 + \ii\, 0.526944$ & $0.24575 + \ii\,0.515117$    & $0.415019 + \ii\, 0.493248$\\
    $+0.9695$  & 0.94496 &$0.054769 + \ii\, 0.736715$  & $0.16459 + \ii \,0.734763$   & $0.275309 + \ii\, 0.730986$ \\
\end{tabular}
  \end{ruledtabular}
\end{center}
\end{table*}

In this paper, we propose a new, alternative, NR-informed strategy for constructing accurate analytical representations 
of the ringdown waveform of coalescing, spinning, black hole binaries. Here we shall consider only nonprecessing, 
equal-mass and equal-spin binaries, but our method is general, being based on a new way of analyzing and 
fitting the ringdown signal provided by NR simulations. In this paper we rely on recent progress in NR simulations and in particular 
on the free availability of hundreds of NR simulations in the Caltech-Cornell 
{\it Simulating eXtreme Spacetimes} (SXS) catalogue~\cite{SXS:catalog,Mroue:2013xna,Lovelace:2010ne}.

The approach pursued here aims at having an {\it effective} and accurate representation of the ringdown. 
Our approach does not aim (contrary to Refs.~\cite{Berti:2007fi,Berti:2007zu,Berti:2007dg,Hadar:2011vj,Taracchini:2014zpa,London:2014cma})
at extracting the actual QNM content of NR ringdown waveforms, nor the excitation coefficients of each mode. 
Our work is similar in spirit to, though technically quite different from, the phenomenological 
ringdown model introduced in Baker et al.~\cite{Baker:2008mj}.

This paper is organized as follows. In Sec.~\ref{sec:data} we briefly review the SXS data that we use. 
Section~\ref{sec:complex} introduces the new tool on which our analysis is based: the QNM-rescaled complex
ringdown waveform $\bar{h}(\tau)$. Section~\ref{sec:fits} describes in detail the fitting procedure 
we applied to each QNM-rescaled ringdown waveform and discusses general fitting formulas that can 
be used outside the set of NR simulations at our disposal. Concluding remarks and outlook are 
collected in Sec.~\ref{sec:conclusion}. We set $G=c=1$.

%-----------------------------------------
\section{Numerical waveform data}
\label{sec:data}
%-----------------------------------------
We use sixteen waveforms produced  by the Caltech-Cornell collaboration with the S$_{\rm P}$EC code. These waveforms 
are publicly available through the SXS catalog~\cite{SXS:catalog} (these data were originally published in 
Refs.~\cite{Chu:2009md,Lovelace:2010ne,Lovelace:2011nu,Buchman:2012dw,Mroue:2012kv,Mroue:2013xna,Hemberger:2013hsa}). 
All waveforms are equal-mass ($m_1=m_2$), equal-spin, with the individual spins either both aligned or
antialigned with the orbital angular momentum.
The dimensionless individual spins are (after relaxation)
\begin{align}
\chi\equiv \chi_{1}=\chi_{2} = &(0.9794, 0.9695, 0.9496, 0.8997,\nonumber\\
                             & 0.8498, 0.7999, 0.6000, 0.43655,\nonumber\\ 
                             & 0.2000, 0, -0.2000, -0.43756,\nonumber\\
                             -&0.5999, -0.7998, -0.8996, -0.9495)\nonumber.
\end{align}
We will simply refer to them as $\chi= (0.98,0.97,\pm0.95, \pm 0.9, 0.85,\pm 0.8,\pm 0.6, \pm0.44,\pm 0.2, 0)$.

We use the highest-resolution waveforms present in the catalogue, extrapolated at future null infinity 
using a 3rd-order polynomial ($N=3$ label in the data). We deal here only with the 
asymptotic\footnote{i.e., rescaled by a factor ${\cal R}/M$.}
$\ell=m=2$ metric waveform
$h_{22}=A_{22}e^{-\ii \phi_{22}}$ (with $\omega_{22}=\dot{\phi}_{22}> 0)$, 
and denote its $\nu$-rescaled version (with $\nu=m_1 m_2/(m_1+m_2)^2$) as $h\equiv h_{22}/\nu$.

Figure~\ref{fig:wave_cmplx} (for the high spin case $\chi=0.9695\approx 0.97$) plots the complex number $h$ versus time, as a curve 
in a 3-dimensional space, focusing on the part of the waveform around its peak. In this paper, we define {\it merger}, occurring 
at $t=t_{0}$, as the {\it peak of the modulus of $h$}. Correspondingly, {\it ringdown} is defined as the signal after merger, $t>t_{0}$:
it is depicted as the thicker (red online) part of the plot.

%==============================================
\section{Complex-number-based approach to QNM generation}
\label{sec:complex}
%==============================================

\begin{figure}[t]
\begin{center}
 \includegraphics[width=0.45\textwidth]{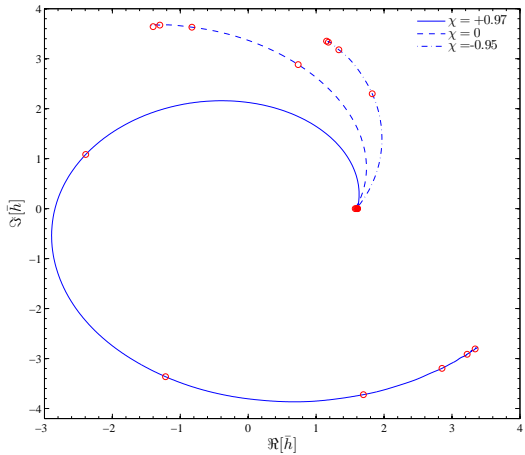}
    \caption{ \label{fig:mrg_vs_chi}Behavior of the QNM-rescaled waveform $\bar{h}$ in the complex plane for three 
                                    values of $\chi$. The filled circle corresponds to merger time $\tau=0$. The empty 
                                    circles mark $\Delta \tau=10$ time intervals. The time extension of each 
                                    curve is $\tau_{\rm max}(\chi)=3.8/\alpha_1(\chi)$, where $\alpha_1(\chi)$ is
                                    the ($M_{\rm BH}$-rescaled) inverse damping time of the first 
                                    QNM as given in Table~\ref{tab:QNMs}.} 
\end{center}
\end{figure}

Figure~\ref{fig:wave_cmplx} highlights the complex-number nature of
the ringdown signal. 
Here we shall show how to get a reliable effective representation of the 
ringdown signal $h(t)$ ($t>t_{0}$) by means of a 
{\it multiplicative decomposition} of the complex number $h(t)=h_1(t)\bar{h}(t)$, 
instead of the usual linear, additive, QNM decomposition $h(t)=h_1(t)+h_2(t)+h_3(t)+\cdots$.
In the following we work with the dimensionless time parameter  $\tau\equiv (t-t_{0})/M_{\rm BH}$
which counts time in units of the final black hole mass $M_{\rm BH}$.
The basic new idea of our approach is to factor out of $h(\tau)$ the contribution 
of the fundamental QNM, $h_1(\tau)\propto \exp[-\sigma_1 \tau]$ , where $\sigma_1=\alpha_1+\ii\omega_1$
is the (dimensionless, $M_{\rm BH}$-rescaled) complex frequency of the fundamental 
(positive frequency, $\omega_1>0$) QNM, by defining the following  {\it QNM-rescaled ringdown waveform}
\be
\label{eq:barh_def}
\bar{h}(\tau) \equiv e^{\sigma_1\tau+\ii\phi_{22}^{\rm mrg}}h(\tau),
\ee
where $\phi_{22}^{\rm mrg}$ is the value of $\phi_{22}$ at merger
(so that $\bar{h}(\tau=0)=A_0$ is the real amplitude of the waveform $h$ at merger). 
In a loose sense we can think of $\bar{h}$ as being the ringdown signal viewed in a frame 
rotating with the complex frequency $\omega_1-\i\alpha_1$.
Fig.~\ref{fig:mrg_vs_chi} plots the parametrized curves drawn by $\bar{h}(\tau)$ in the complex plane 
for three values of $\chi=(-0.94905,0,+0.9695)\approx(-0.95,0,+0.97)$. The filled circle corresponds to the beginning of the
ringdown, $\tau=0$. Note that the modulus of the waveform at merger is nearly independent
of $\chi$~\cite{Taracchini:2012ig}, so that the three curves start nearly at the same 
point ($\bar{h}(0)\approx 1.59$)\footnote{The small variations with $\chi$ of the merger amplitude 
$A_0\equiv A_{22}^{\rm mrg}/\nu$ will be quantified below.} 
The other, empty, circles mark time intervals of 
$10M_{\rm BH}$ after merger. We have stopped the three curves at the  
$\chi$-dependent time $\tau_{\rm max}(\chi)=3.8/\alpha_1(\chi)$, corresponding to a 
fixed decrease in the modulus of the first QNM by 
a factor $\exp(-3.8)\approx 1/44.7$.

%=================================
% Fig. 2
%=================================
%----------------------------------------------------
%=================================
% Fig. 3
%=================================
\begin{figure}[t]
\begin{center}
 \includegraphics[width=0.45\textwidth]{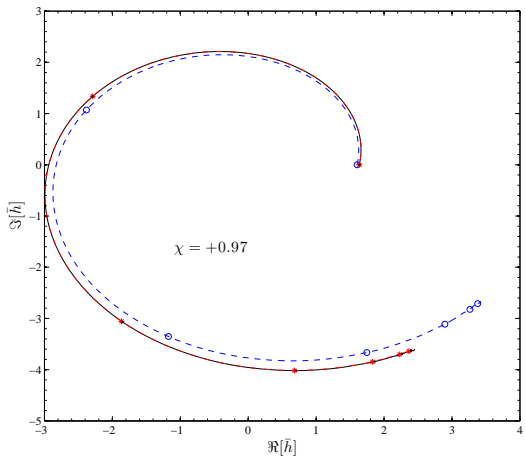}
    \caption{ \label{fig:barh_error}The $\chi=+0.97$ complex-plane QNM-rescaled waveform $\bar{h}$ 
    for different resolutions (either Level~5 or Level~6) and radius-extrapolations (either $N=3$ or $N=4$). 
    Shown are: (i) the highest resolution 
    (Level~6, black line with time-markers indicated as dots) with $N=3$; (ii) the medium resolution 
    (Level~5, red line with time-markers indicated as stars) with $N=3$;
    and (iii) the $N=4$ radius-extrapolation of the high-resolution $\bar{h}$ 
    (dashed line with time-markers as circles).} 
\end{center}
\end{figure}
%----------------------------------------------------
The striking change in the shape of these curves as $\chi$ increases from $\chi=-0.95$ to
$\chi=+0.97$ is a new window on the process of QNM generation after merger.
The expected analytical expression of $\bar{h}$ under the usual assumption that the 
QNMs are already generated just after merger would be
\be
\label{eq:QNM_anlyt}
\bar{h}_{\rm QNM}(\tau)= c_1 + c_2 e^{-(\sigma_2-\sigma_1)\tau}+c_3 e^{-(\sigma_3-\sigma_1)\tau}+ \cdots ,
\ee
with some {\it constant}, complex coefficients $c_i$.
In Table~\ref{tab:QNMs} we list, for the three curves, the values of the dimensionless 
spin parameter $\hat{a}_{\rm BH}=J_{\rm BH}/M_{\rm BH}^2$ of the final black hole as 
well as the complex frequencies of the corresponding first three QNMs. 
Note that, while the damping coefficients $\alpha_i=\Re[\sigma_i]$ of the successive QNMs
increase with QNM order (with only the first one $\alpha_1$ being smallish compared to 1), 
the real frequencies $\omega_i$'s are nearly independent of QNM order, especially when
the spin of the final black hole gets large\footnote{Reference~\cite{Onozawa:1996ux} 
found this high-spin behavior to hold for the first 20 overtones {\it except} for the sixth one whose
real frequency does not cluster with the others. The presence of such a single anomalous overtone 
does not significantly affect our conclusions below.}. 
Therefore, the complex frequency differences $\sigma_{21}\equiv \sigma_2-\sigma_1$, 
$\sigma_{31}\equiv \sigma_3-\sigma_1$ entering Eq.~\eqref{eq:QNM_anlyt} are approximately
real and positive, especially for high spin, e.g. for $\chi=0.9695\approx 0.97$ one has 
$\sigma_{21}\approx 0.109821 - \ii\,0.001952\approx 0.11$.
As a consequence the curve parametrized by the mathematical expression Eq.~\eqref{eq:QNM_anlyt} 
describes an {\it approximately straight line} in the complex plane of $\bar{h}$.
The curve in Fig~\ref{fig:mrg_vs_chi} corresponding to $\chi=-0.95$ is approximately
straight and therefore can be well represented by a superposition of QNMs as in
Eq.~\eqref{eq:QNM_anlyt} starting at merger; i.e., for $\tau\geq 0$.
The $\chi=0$ curve is still approximately straight, so that it can also be represented
by a QNM sum, if one includes sufficiently many modes\footnote{Previous EOB works used
either $N=5$ modes~\cite{Damour:2012ky} or $N=8$ modes~\cite{Pan:2011gk} to reliably 
match the ringdown waveform.}.
By contrast, the $\chi=+0.97$ case leads to a snail-shaped curve, which cannot be (easily) 
globally represented by a QNM sum of the type~\eqref{eq:QNM_anlyt}. We interpret this as
a hint that the QNM generation is not yet completed at merger so that one would probably 
need to use {\it time-dependent} coefficients $c_i(\tau)$ in Eq.~\eqref{eq:QNM_anlyt} to 
represent the ringdown signal by a (short) QNM sum.
On the other hand, we see (still for $\chi=+0.97$) that after $\tau=20$ the $\bar{h}$ curve
is approximately straight so that for $\tau\geq 20$ Eq.~\eqref{eq:QNM_anlyt} (with constant
coefficients) would allow for a reliable representation of the ringdown signal.
One can say that, for $\chi=+0.97$, the generation of QNMs is completed only $\sim 20M_{\rm BH}$ 
after merger.
In Fig.~\ref{fig:barh_error} we investigate the robustness of the complex-plane $\bar{h}$ 
behavior under changes of resolutions and/or radius-extrapolation order. Changing the resolution
has nearly no effect (apart from slightly displacing the $\Delta \tau=10$ time markers), while
the effect of radius-extrapolation is more significant. In keeping with the results of 
Ref.~\cite{Boyle:2009vi} we use in the present work the $N=3$ extrapolated data as a compromise.

The complex-plane, $\bar{h}$, representation illustrated in Fig.~\ref{fig:mrg_vs_chi}
gives a new understanding of the practical need, in case one insists on representing 
the ringdown signal as a sum of complex frequency modes, to go beyond the actual QNM frequencies 
$\sigma_i=\alpha_i+\ii\, \omega_i$ by including pseudo-QNMs frequencies 
$\sigma_i'=\alpha_i'+\ii\, \omega_i'$~\cite{Pan:2011gk,Taracchini:2012ig,Taracchini:2013rva}, 
whose role is to account for the rotation (with sizable nonzero real frequency difference, 
say $\omega_2'-\omega_1$)  we see in the $\chi=0.97$ curve. [We have checked that a similar
$\bar{h}$-plane behavior explains the need for pseudo-QNM frequencies for higher multipolar 
waveforms, e.g. $\ell=m=4$, even in the case of low or negative $\chi$].

%=================================
% Fig. 3
%=================================
\begin{figure}[t]
\begin{center}
 \includegraphics[width=0.45\textwidth]{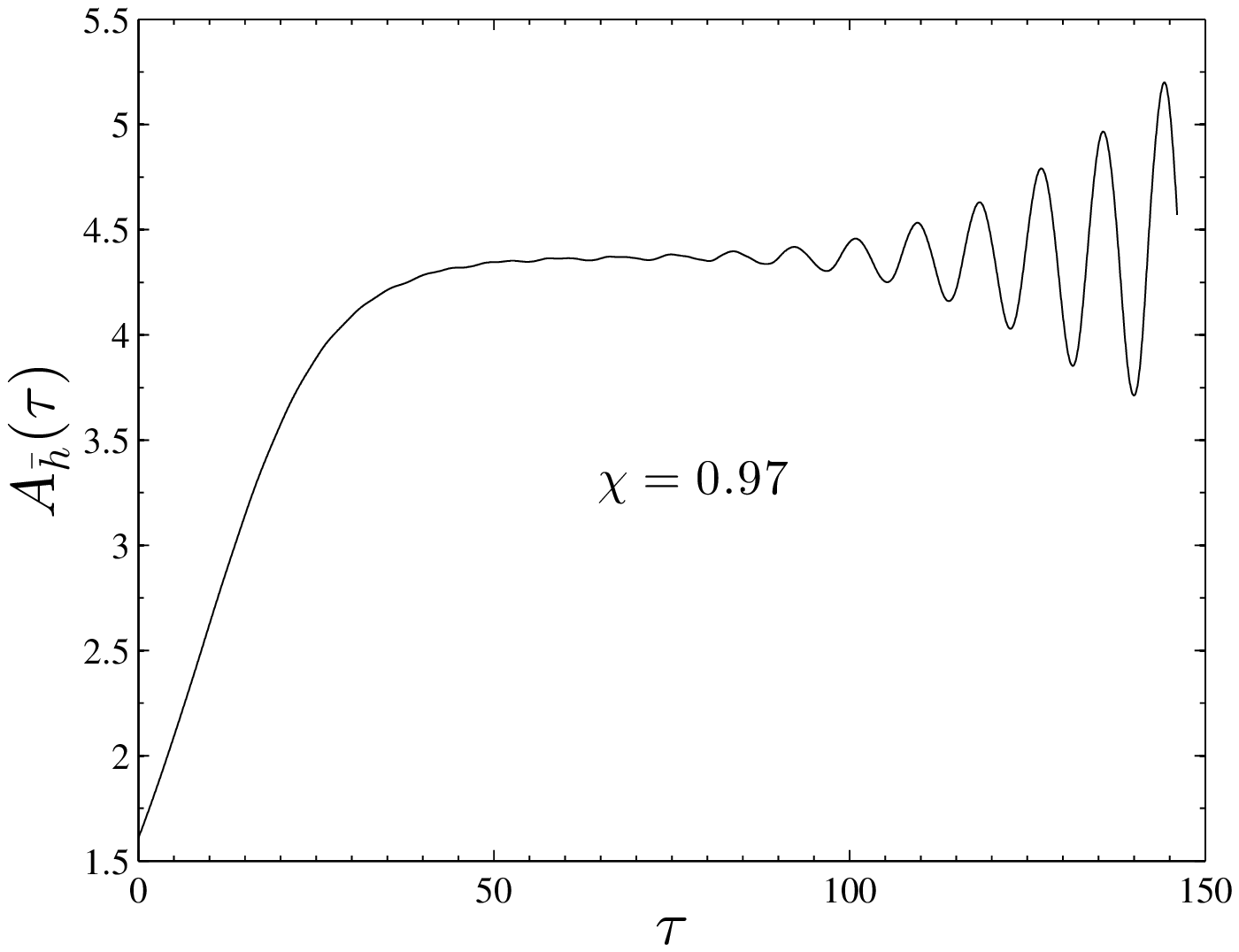}
 \includegraphics[width=0.45\textwidth]{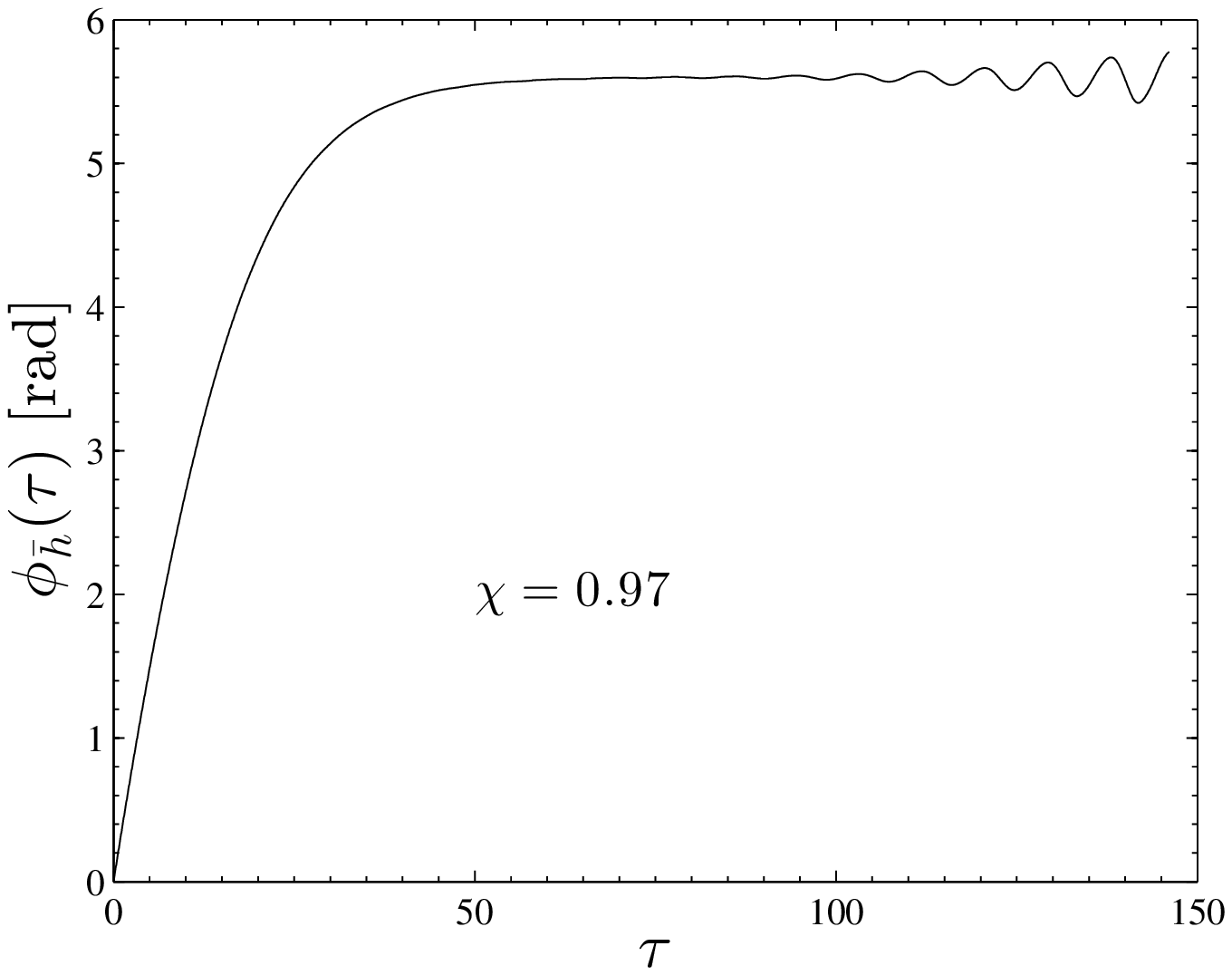}
    \caption{ \label{fig:Aphi_barh}Time evolution of the amplitude and phase of $\bar{h}$ for $\chi=0.97$.
               This plot is more extended in time than the corresponding complex-plane representation
               of $\bar{h}$ in Fig.~\ref{fig:mrg_vs_chi} to highlight the amplification
               of numerical noise for very late times.} 
\end{center}
\end{figure}

%=================================
% Fig. 4
%=================================
\begin{figure}[t]
\begin{center}
 \includegraphics[width=0.45\textwidth]{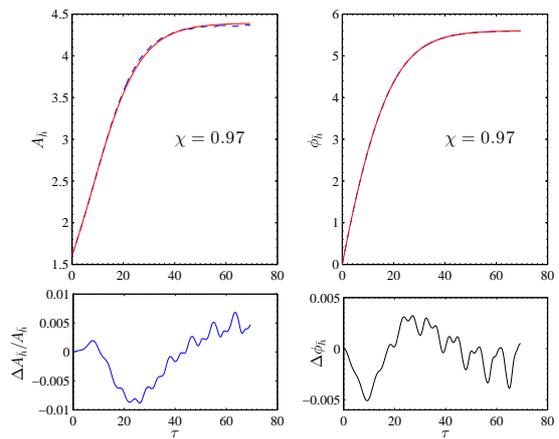}
 \caption{ \label{fig:A_and_phi}Top panels: quality of the fits of $A_\h$ and $\phi_\h$ for  $\chi=0.97$. The 
   residuals are shown in the bottom panels.} 
\end{center}
\end{figure}

%===================================
% Fig. 5
%===================================
\begin{figure*}[t]
\begin{center}
 \includegraphics[width=0.45\textwidth]{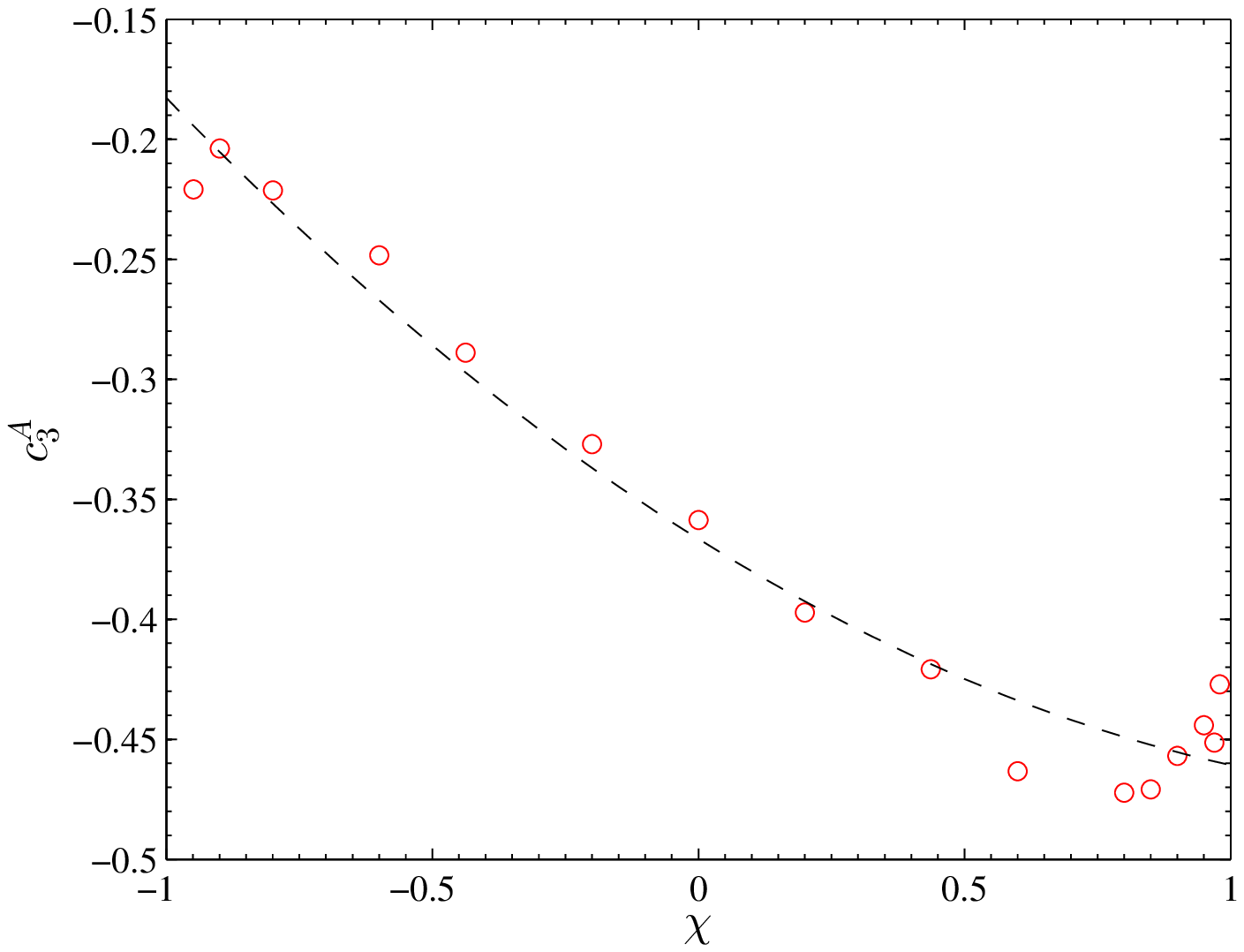}
 \includegraphics[width=0.45\textwidth]{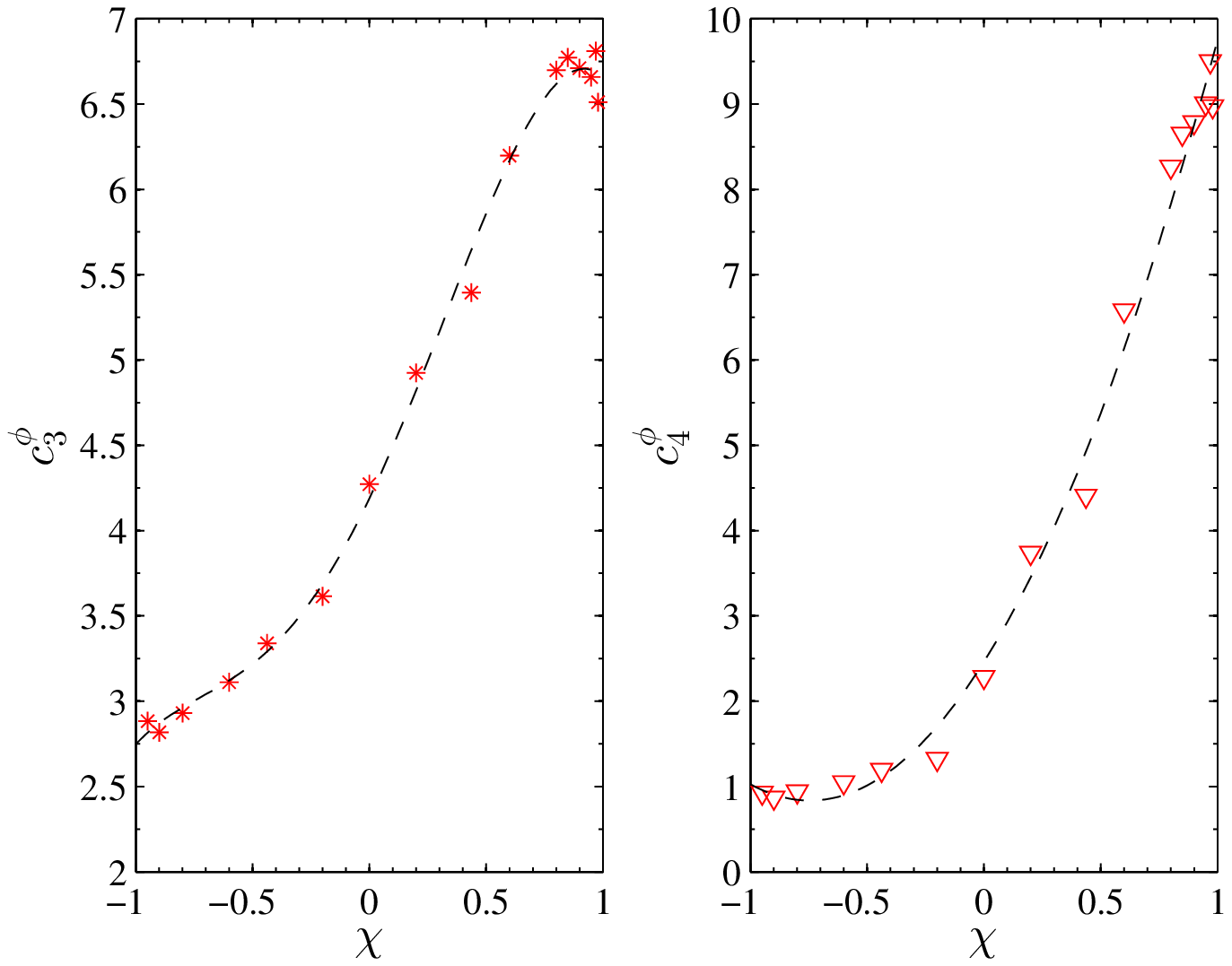}
    \caption{ \label{fig:coeffs}Behavior of the six fitting coefficients versus dimensionless spin $\chi$. The 
    numerical points can be well fitted by simple polynomials in $\chi$. The corresponding coefficients 
    are listed in Table~\ref{tab:coeff_info}.} 
\end{center}
\end{figure*}
%=================
% Table II
%=================
\begin{table}[t]  
\caption{\label{tab:coeff_info} Coefficients of the polynomial representations of: (i) the $(c_3^A; c_3^\phi,\,c_4^\phi)$ 
coefficients entering the amplitude and phase fitting templates~\eqref{Atemp}-\eqref{phitemp}, and (ii) the QNM/NR data
entering the osculation constraints Eqs.~\eqref{const1}-\eqref{const4}, with $\Delta\omega\equiv\omega_{1}-M_{\rm BH}\omega_{22}^{\rm mrg}$.}
  \begin{center}
    \begin{ruledtabular}
\begin{tabular}[c]{llllll}
       &  $p_4$  & $p_3$  &  $p_2$           &      $p_1$          &    $p_0$  \\
\hline 
$c_3^A$                & 0         & 0          & 0.044763   & $-0.138980$ & $-0.366538$ \\
$c_3^\phi$             & $-1.174263$    & $-0.9099211$    & 1.690678  & 2.866629 & 4.188784 \\
$c_4^\phi$              & 0         & 0          & 2.925663   & $4.362706$& 2.462696 \\
\hline
$\hat{A}^{\rm mrg}_{22}$  & $0.014175$  & 0.014553   & 0.012896   & $-0.004458$ & 1.575613 \\
$\alpha_{21}$           & $-0.009068$ & $-0.013719$  & $-0.012981$  & $-0.022385$ & 0.164398 \\
$\alpha_1$             & $-0.004416$ & $-0.006810$ & $-0.006789$  & $-0.010196$ & $0.081224 $\\
$\Delta\omega$         & $0.020975$  & 0.028444   & 0.026957   & $0.066588$ & 0.184738 \\
\end{tabular}
  \end{ruledtabular}
\end{center}
\end{table}

%=====================================================
\section{Amplitude-phase approach to ringdown fitting}
\label{sec:fits}
%==============================================
%=================================
% Fig. 6
%=================================
\begin{figure}[t]
\begin{center}
 \includegraphics[width=0.42\textwidth]{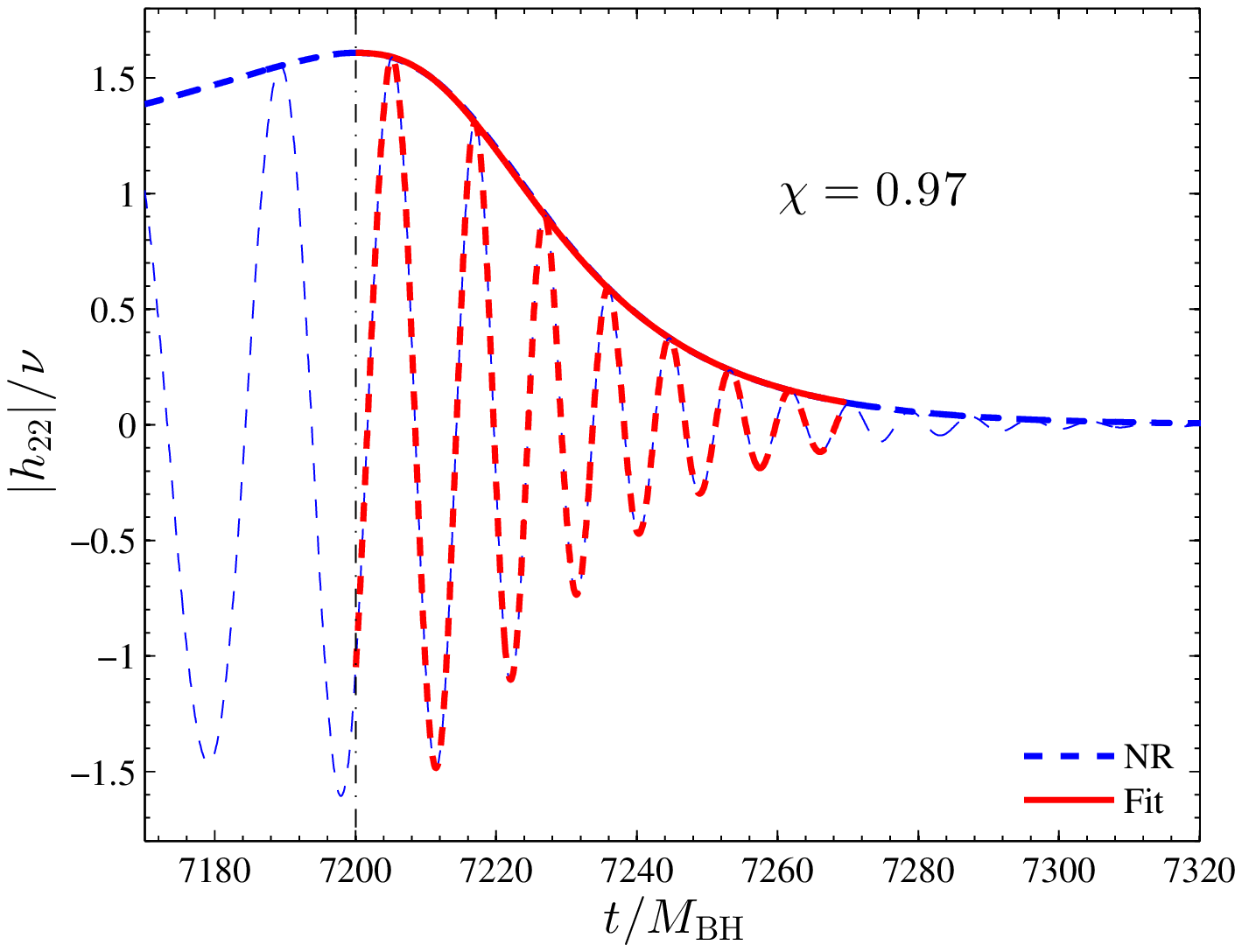}\\
 \includegraphics[width=0.42\textwidth]{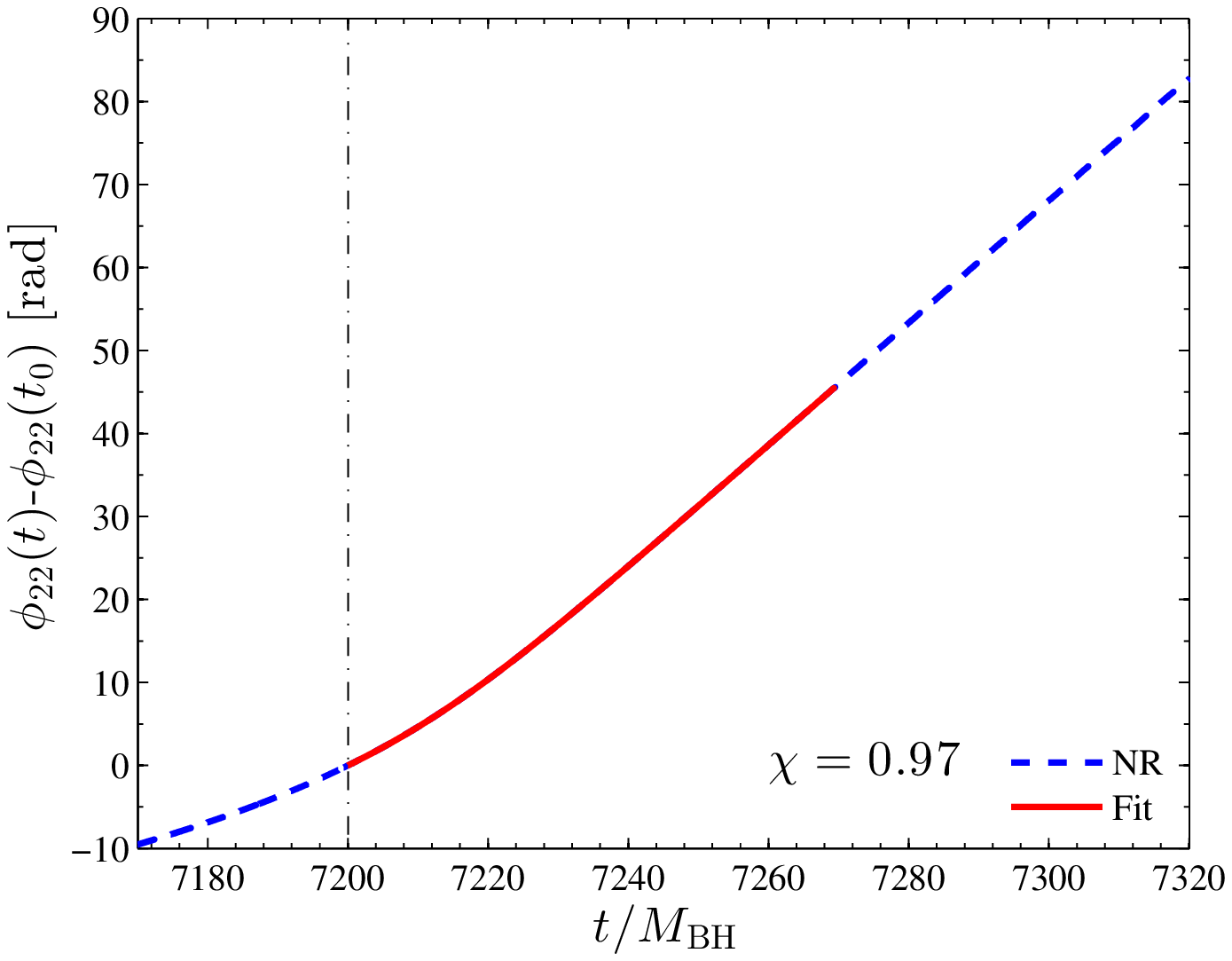}\\
 \includegraphics[width=0.42\textwidth]{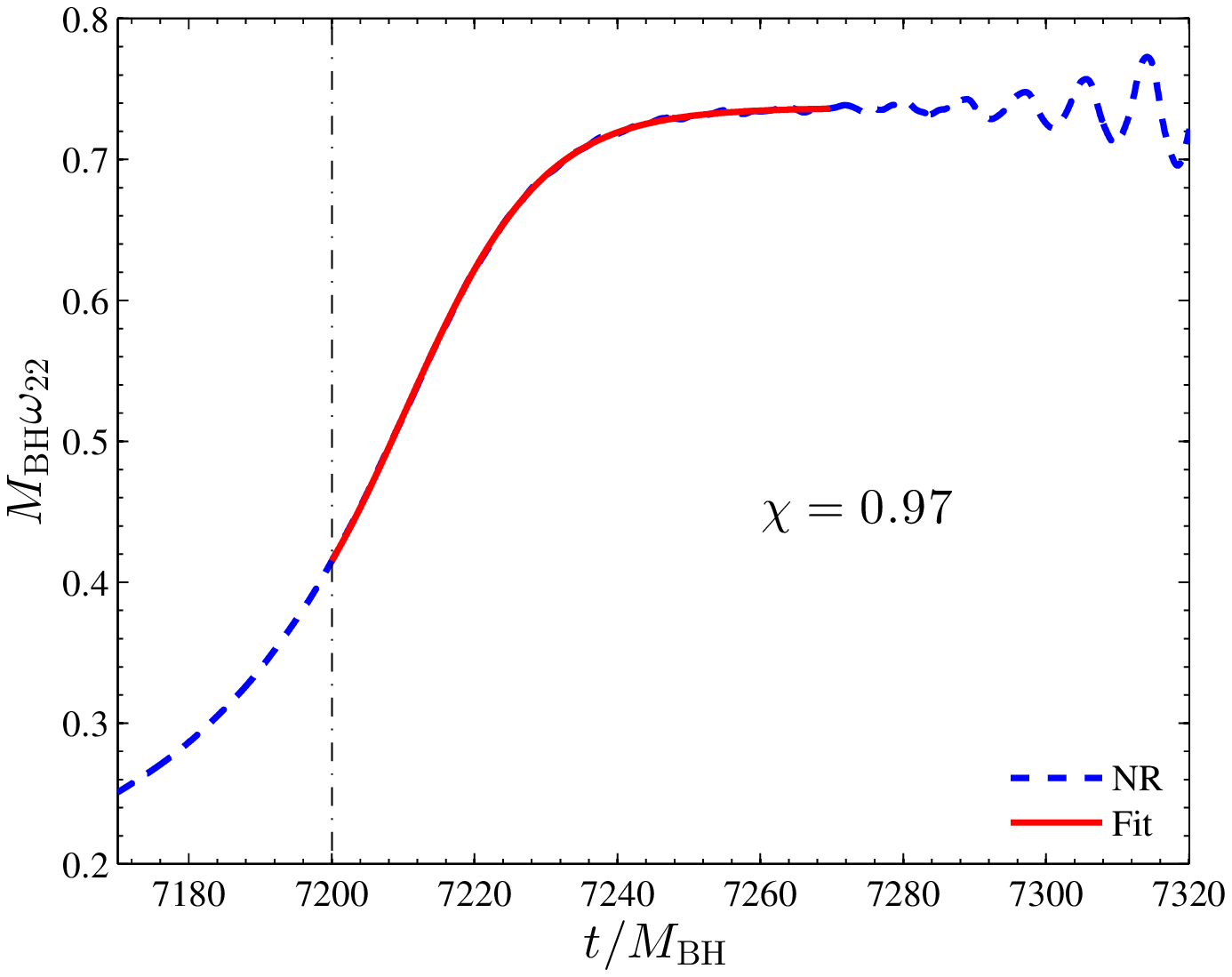}
    \caption{ \label{fig:freq_amp}NR waveform (dashed line) and its ringdown fit (solid line, for $0\leq \tau \leq 3.8/\alpha_1(\chi)$) 
              for $\chi=0.97$: amplitude (top panel), phase (middle panel) and frequency (bottom panel). The vertical lines correspond 
              to the merger, defined as the peak of $|h_{22}|/\nu$.} 
\end{center}
\end{figure}
If one wanted to represent the ringdown signal as a short sum of complex frequency modes,
the data behind the $\bar{h}$ curves of Fig.~\ref{fig:mrg_vs_chi} could be used to 
find optimal values of any additional pseudo-QNM frequency $\sigma'_i$.
However, we wish here to suggest an alternative strategy for analytically representing the
ringdown signal.
The QNM-rescaled ringdown waveform $\bar{h}(\tau)\equiv e^{\sigma_1\tau+\ii\phi_{22}^{\rm mrg}}h(\tau)$ 
being a complex quantity, can be decomposed in
amplitude and phase
\be
\bar{h}(\tau)\equiv A_{\h}(\tau) e^{+\ii \phi_\h(\tau)}.
\ee
[Note the sign convention: $\phi_{\h}(\tau)=\omega_1\tau - \phi_{22}(\tau)+\phi_{22}^{\rm mrg}$].
Figure~\ref{fig:Aphi_barh} exhibits the time evolution (after merger $\tau\geq 0$) of
the amplitude and phase $(A_\h,\phi_\h)$ of $\bar{h}$ for $\chi=0.97$.
The shapes of the curves $A_\h(\tau)$ and $\phi_\h(\tau)$ are similar and are both reminiscent
of a hyperbolic tangent. Other values of $\chi$ lead to very similar shapes.
The plateau behavior of the curves as $\tau$ increases is linked to the asymptotic behavior 
$\bar{h}(\tau)\approx \bar{h}_{\rm QNM}(\tau)\sim c_1 + {\cal O}\left(e^{-(\alpha_2-\alpha_1)\tau}\right)$ 
given by Eq.~\eqref{eq:QNM_anlyt}.
[The oscillations around the plateau that one sees on these curves for $\tau\gtrsim 80$
are due to the amplification of numerical noise by the exponentially growing
factor $e^{\sigma_1\tau}$].
We found that, at an effective level,  $A_\h(\tau)$ and $\phi_\h(\tau)$ can be 
accurately represented by the following general functional forms
\begin{align}
\label{Atemp}
A_\h(\tau)    & = c_1^A \tanh\left(c_2^A \tau + c_3^A\right) + c_4^A,\\
\label{phitemp}
\phi_\h(\tau) & = -c_1^\phi\ln\left(\dfrac{1+c_3^\phi e^{-c_2^\phi\tau}+c_4^\phi e^{-2c_2^\phi\tau}}{1+c_3^\phi+c_4^\phi}\right).
\end{align}
[Note that when $c_4^\phi=0$, the time-derivative of $\phi_\h(\tau)$ differs by a constant 
from a hyperbolic tangent].
In addition, we constrain the parameters entering these functional forms by imposing
simple physical constraints similar to the ones imposed in the usual linear QNM
representation of the ringdown. Namely, we impose: (i) that the value of $A_\h(\tau)$
at $\tau=0$ coincide with the NR amplitude at merger $\hat{A}_{22}^{\rm mrg}\equiv A_{22}^{\rm mrg}/\nu$; 
(ii) the value of $dA_\h(\tau)/d\tau$ at $\tau=0$ so that the $\dot{A}_{22}=0$ 
at merger; (iii) that the amplitude exponential-decay parameter $c_2^A$ coincide with $\alpha_{21}/2$, 
where $\alpha_{21}\equiv \alpha_2-\alpha_1$; (iv) that $d\phi_\h(\tau)/d\tau$ 
at $\tau=0$ be equal to $\Delta\omega\equiv \omega_1-M_{\rm BH}\omega_{22}^{\rm mrg}$; 
and (v) that the phase exponential-decay parameter $c_2^\phi$ be equal to $\alpha_{21}$.
Imposing these five constraints allow one to keep only {\it one free parameter}, 
namely $c_3^A$, in the amplitude template Eq.~\eqref{Atemp}, and {\it two free
parameters}, namely $c_3^\phi$ and $c_4^\phi$, in the phasing template, Eq.~\eqref{phitemp}.
Specifically, the other parameters are expressed in terms of these three 
independent parameters as follows
\begin{align}
\label{const1}
c_2^A   & = \frac12 \alpha_{21},\\
c_1^A   & = \hat{A}_{22}^{\rm mrg}\alpha_1\dfrac{\cosh^2 c_3^A}{c_2^A},\\
c_4^A   & = \hat{A}_{22}^{\rm mrg}-c_1^A\tanh c_3^A,\\
\label{const3}
c_2^\phi &= \alpha_{21},\\
\label{const4}
c_1^\phi &= \dfrac{1 + c_3^\phi + c_4^\phi}{c_2^\phi(c_3^\phi + 2 c_4^\phi)}\left(\omega_1-M_{\rm BH}\omega_{22}^{\rm mrg}\right).
\end{align}
In the present paper, we propose as effective strategy for analytically representing the 
ringdown to least-square fit the QNM-rescaled NR ringdown waveform, $\bar{h}_{\rm NR}(\tau)$,
to the templates~\eqref{Atemp},\eqref{phitemp}, constrained by 
Eqs.~\eqref{const1}-\eqref{const4}, so as to determine best-fit values of 
the three coefficients  $c_3^{A}$, $c_3^\phi$ and $c_4^\phi$.
We have chosen as $\tau$ interval for the fitting $0\leq \tau \leq 3.8/\alpha_1(\chi)$.
Indeed  $\tau_{\rm max}=3.8/\alpha_1(\chi)$ happens to be well on the plateau while still
avoiding the region where the oscillations get significant (e.g., $\tau_{\rm max}\approx 69.38$ for $\chi=0.97$).
The quality of the fit performance is illustrated in Fig.~\ref{fig:A_and_phi} for $\chi=0.97$. 
Note in particular how the phasing is reproduced within $5\times 10^{-3}$ radians.
In a first version of this analysis we used the same (but unconstrained) amplitude template 
and the following simpler, two-parameter, unconstrained phasing template 
$ \phi_\h(\tau)  = c_1^\phi \tanh\left(c_2^\phi \tau\right)$. Such choices led to a comparably
accurate representation of the ringdown for $\tau\gtrsim 10$, but to larger disagreements 
for $0\leq \tau \lesssim 10$.

We have applied this strategy to the sixteen waveforms of the SXS catalog. For each waveform, 
i.e. for each $\chi$, the ringdown information is encoded in the set of three coefficients 
$(c_3^A;\,c_3^\phi,c_4^\phi)$.
We have found that the $\chi$-dependence of the $c_i^{(A,\phi)}$'s is relatively smooth (see Fig.~\ref{fig:coeffs}).
One can approximately represent these coefficients as second-order polynomials in $\chi$
except for $c_3^\phi$ for which we found that a fourth-order polynomial gives a better 
fit\footnote{For completeness, let us mention that a (less accurate) second-order fit for 
$c_3^\phi$ reads $c_3^\phi=0.479448\chi^2+2.176818\chi+4.342270$.}. 
To complete the information needed to use our results we also provide fits 
for the $\chi$-dependence of $\alpha_1$, $\alpha_{21}$, $\hat{A}_{22}^{\rm mrg}$ 
and $\Delta\omega\equiv \omega_1 - M_{\rm BH}\omega_{22}^{\rm mrg}$.
All our fits are done with the convention 
$c_i(\chi) = p_{4}\chi^{4} + p_{3}\chi^{3} + p_{2}\chi^{2} + p_{1}\chi + p_{0}$.
The explicit values of the $p_n$ coefficients are listed in Table~\ref{tab:coeff_info}. 

The comparison between the $\chi$-fits for $(c_3^A;\,c_3^\phi,c_4^\phi)$ and the raw points is 
displayed in Fig.~\ref{fig:coeffs}. The amplitude coefficient plot shows more scatter around the fit, probably 
due to amplified numerical noise (this is consistent with the fact that the oscillation around the plateau
is larger for amplitude than for phase, see Fig.~\ref{fig:Aphi_barh}). We have checked that by reducing 
the $\tau$-length of the fitting interval the scatter could be reduced, especially for large spins.
Changing the extrapolation order $N=3$ to $N=2$ reduces the oscillations around the plateaux in 
Fig.~\ref{fig:Aphi_barh} and thereby the scatter. We have checked that inserting the $\chi$-fitted  versions 
of $(c_3^A;\,c_3^\phi,c_4^\phi)$ and of $(\hat{A}_{22}^{\rm mrg},\;\alpha_{21},\;\alpha_1,\Delta\omega)$
in our functional forms Eq.~\eqref{Atemp}-\eqref{const4} leads to representations of the ringdown
with phase and fractional amplitude disagreements that remain smaller than about 0.04 in all cases.

Finally, the very satisfactory representation, given by our strategy, of the original
ringdown waveform $h(t)=h_{22}/\nu$ (decomposed in amplitude, phase and frequency), is displayed in 
Fig.~\ref{fig:freq_amp} for the case $\chi=0.97$. The corresponding phase and fractional amplitude 
differences were given in the bottom panels of Fig.~\ref{fig:A_and_phi}.

%================================
\section{Conclusions and outlook}
\label{sec:conclusion}
%================
Let us summarize our main results.
\begin{enumerate}
\item We introduced a new tool for analyzing ringdown waveforms, consisting of studying the
time evolution, after merger, of the QNM-rescaled complex quantity $\bar{h}(\tau)$, Eq.~\eqref{eq:barh_def}.
\item Using publicly available SXS, Caltech-Cornell, waveform data~\cite{SXS:catalog} spanning
dimensionless spins $-0.95\leq \chi \leq 0.98$, and the $\bar{h}$ tool, we have found that, in the case 
of large spin, the ringdown signal is compatible with the usually expected
sum, Eq.~\eqref{eq:QNM_anlyt}, of Kerr black hole quasi-normal-modes complex frequencies, {\it only} for
times sufficiently posterior to merger, e.g., by about $20M_{\rm BH}$ when $\chi=0.97$.
For earlier times the $\bar{h}$ diagnostics can be seen as a new tool to study the building up of QNMs just
after merger.
\item To get an analytic representation of the ringdown signal starting just after merger we emphasized 
that two strategies are possible: (i) to introduce, as 
in Refs.~\cite{Pan:2011gk,Taracchini:2012ig,Taracchini:2013rva}, pseudo-QNM frequencies, in which
case our $\bar{h}$ diagnostics can provide an efficient tool for optimizing their determination;
or (ii) to separately fit the amplitude and phase of the QNM-rescaled signal $\bar{h}(\tau)$ by
hyperbolic-tangent-based templates.
\item We have applied the latter strategy to sixteen, equal-mass, equal-spin, $\ell=m=2$ waveforms 
of the SXS catalog and showed that it leads to a very accurate representation of the ringdown waveform 
(with phase differences comparable to numerical errors). We have checked that our strategy also yields
accurate representations of ringdown waveforms for {\it unequal mass}, spin-aligned SXS waveforms.
\item We provided explicit polynomial representations as functions of $\chi$ of the coefficients entering 
our fitting templates for the amplitude and phase of the ringdown signal. 
\item
The quality of our fits suggests that our method will give reliable representations of the ringdown
waveforms also for values of the spin not included in the SXS catalog. The method allows one to interpolate between
the catalogued $\chi$ values, and hopefully also to {\it extrapolate} the full ringdown waveform to more 
extremal spin values. For example, we can predict the energy radiated during the ringdown versus spin.

\end{enumerate}

Our findings open the following avenues for further research.
\begin{itemize}
\item[]{(a)} The extension of our approach to higher multipolar modes ($\ell>2$) is conceptually straightforward, 
             but should be quantified. In particular, it will be interesting to investigate to what extent the
             behavior of higher multipolar modes during ringdown {\it is not} representable [because of an excessive
             rotation of $\bar{h}(\tau)$] as a sum of QNMs with constant coefficients.

\item[]{(b)} A preliminary investigation of the behavior of $\bar{h}(\tau)$ in the ringdown waveform generated 
             by a point particle inspiralling and plunging on a (fast-spinning) Kerr black hole has shown that the 
             rotating features of $\bar{h}(\tau)$ are even more marked in that case~\cite{Nagar:prep}. Future work 
             on this case will hopefully improve our knowledge of the generation mechanism of QNMs.

\item[]{(c)} The optimal choice of extrapolation order for the numerical waveform must be further investigated.
             Similarly, it will be interesting to analyze also different NR waveform data obtained with independent 
             infrastructures, so as to gauge possible (tiny) systematics present in extrapolated SXS data.

\item[]{(d)} The quality of our fits should be quantified, using data-analysis-relevant measures. This might allow
             one to use variations on our fits such as: (i) deleting some of our constraints; (ii) modifying 
             them\footnote{For instance by replacing the second exponential $e^{-2c_2^\phi\tau}$ in Eq.~\eqref{phitemp}
             by $e^{-(\alpha_3-\alpha_1)\tau}$.}; or (iii) adding further constraints so as to impose a  
             higher order osculation of frequency and amplitude at merger. 
             For instance, we found that relaxing the phase constraints Eqs.~\eqref{const3}-\eqref{const4} leads
             to an even better phasing agreement with a very flat behavior (within $\pm 10^{-3}$~rad) 
             of $\Delta\phi_\h$ as a function of $\tau$.

\item[]{(e)} The fitting procedure presented here could be systematically applied to all waveforms present
             in the SXS catalog, starting from unequal-mass, but spin aligned configurations. 
             Its generalization to non--spin--aligned binaries should be explored.
             Including also higher multipoles and small-mass-ratio waveforms computed from perturbative 
             calculations (solving either the Regge-Wheeler-Zerilli or Teukolsky equations) the procedure 
             outlined here might give an efficient representation of the complete ringdown waveform as a 
             function of the spin magnitude and mass ratio.

\item[]{(f)} Finally, technical ways of using the dynamics of the QNM-rescaled waveform $\bar{h}(\tau)$ for finding
             optimal values of additional pseudo-QNM complex frequencies could be explored and compared to the
             result of the new strategy that we have proposed here.
\end{itemize}

\acknowledgments
We thank Andrea Taracchini for useful comments.

%----------------
% bibliograph
%----------------
\bibliographystyle{apsrev}     
\bibliography{refs20140704.bib}{}

\end{document}